\documentclass[12pt]{article}

\usepackage{authblk}
\usepackage{graphicx}
\usepackage{subfig}
\usepackage{subcaption} 
\usepackage{amsmath}
\usepackage{epstopdf} 
\usepackage{float}
\usepackage{titlesec} 

\def\eq#1{Eq.~(\ref{#1})} 
\def\eqs#1{Eqs.~(\ref{#1})} 
\def\eqn#1{(\ref{#1})} 

\graphicspath{{./Figs/}}

\begin{document}
	
	\bibliographystyle{unsrt}

\title{Single MRT-featured Lattice Boltzmann Method (SmrtLBM)}


\author{Jian Guo Zhou}

\affil{Department of Computing and Mathematics\\
	Manchester Metropolitan University\\
	Manchester, M1 5GD, UK\\
	J.Zhou@mmu.ac.uk}

\date{}

\maketitle


\begin{abstract}
A novel single MRT-featured lattice Boltzmann method (SmrtLBM) is introduced. The intricate algorithm of the multiple-relaxation-time (MRT) collision operator is reimagined and distilled into a streamlined, single-relaxation-time-like scheme. This innovation preserves the simplicity and efficiency of the conventional single-relaxation-time approach while inheriting the enhanced stability characteristic of the MRT techniques, making it a powerful tool for simulating complex fluid flows. Rigorous numerical testing on cavity flows demonstrates that SmrtLBM accurately captures complex flow characteristics, even in challenging high Reynolds number regimes where precision and stability are crucial. Compared to the traditional single-relaxation-time lattice Boltzmann method, this model offers superior stability and accurate solutions for difficult flows.  The SmrtLBM marks a breakthrough in complex fluid simulation through a simple method, offering a new level of efficiency and reliability in tackling challenging flow problems.
\end{abstract}

Among the various lattice Boltzmann method (LBM) variants, the single relaxation time (SRT) model is particularly notable for its efficiency and simplicity, making it one of the most widely used approaches in practical applications. This method exemplifies the core strengths of the LBM framework, including ease of implementation, natural parallelism, and a straightforward treatment of boundary conditions  \cite{ChenDoolen:1998}.  Its potential power is much beyond the original scope, being explored and demonstrated in various disciplines of science and engineering with time \cite{Succi:2001,Wolf:2000,Guo-Shu:2013,Kruger.etc:2017}.
The single relaxation time lattice Boltzmann method (SRT-LBM) reads, 
\begin{equation}
f_\alpha(x_j + e_{\alpha j} \delta t, t + \delta t) 
= f_\alpha(x_j, t)  -
\frac{1}{\tau}  [f_\alpha(x_j , t)-f_\alpha^{eq}(x_j, t)] ,
\label{lb.srt} \end{equation}
where  $f_\alpha$ is the distribution function of particles; $x_j$ is a lattice coordinate along the $j$-axis in the Cartesian coordinate system, e.g., $j=x,\ y$ in the two dimensional space; $t$ represents time; $e_{\alpha j}$ is the $j^{th}$ component of the particle velocity vector ${\bf e}_\alpha$ in $\alpha-$link of the lattice, i.e., ${\bf e}_\alpha = (0,0),\ (e,0),\ (0,e),\  (-e,0),\ (0,-e), \ (e,e),\ (-e,e),\ (-e,-e),\ (e,-e)$ when $\alpha =0 - 8$ for nine particles moving in the two dimensional uniform square lattice (D2Q9), in which $e$ is the particle speed, defined by time step $\delta t$ and lattice size $\delta x$, as $e=\delta x/\delta t$;  $\tau$ is the single relaxation time \cite{Bhatnagar.etc:1954}; and $f_\alpha^{eq}$ is the local equilibrium distribution function given by
\begin{equation}
f_\alpha^{eq} = 
w_\alpha \rho \left( 1+3 \frac{e_{\alpha i}u_i}{e^2}
+ \frac{9}{2} \frac{e_{\alpha i}e_{\alpha j}u_i u_j}{e^4}
- \frac{3}{2} \frac{u_i u_i}{e^2}  \right),
\label{feq-full}
\end{equation} 
here, $\rho$ represents the fluid density, and $w_\alpha$ is a weighting factor that depends on the lattice pattern, e.g., $w_\alpha = 4/9$ when $\alpha = 0$, $w_\alpha = 1/9$ when $\alpha = 1 - 4$ and $w_\alpha = 1/36$ when $\alpha = 5 - 8$ on D2Q9. After the distribution function is calculated from the lattice Boltzmann equation \eqn{lb.srt}, the macroscopic physical variables, density and velocity are calculated as
\begin{equation}
\rho (x_j, t) =\sum_\alpha f_\alpha  (x_j, t), \hspace{13mm}
u_i(x_j, t) = \frac{1}{\rho(x_j, t)} \sum_\alpha e_{\alpha i} f_\alpha (x_j, t).
\label{fea-0}
\end{equation}

The single-relaxation-time lattice Boltzmann method (SRT-LBM) is known to suffer from numerical instability and inaccuracy. Over the years, various approaches have been developed to address these issues. One significant advancement was the introduction of the multiple-relaxation-time (MRT) collision operator in 1992 \cite{dHumieres:1992, Lallemand.etc:2000}, which improved both stability and accuracy, albeit at the cost of computational efficiency. Geier et al. \cite{Geier.etc:2006} further contributed by proposing a cascaded lattice Boltzmann method (CasLBM), which stabilises the LBM for low-viscosity flows by relaxing the particle distribution function towards its local equilibrium state in the central moment space. This approach was later improved in 2015, when Geier et al. \cite{Geier.etc:2015} introduced the cumulant lattice Boltzmann method (CumLBM), using the cumulant in the collision operator to further enhance stability. 
However, as these methods evolve—from MRT to CasLBM to CumLBM—they become increasingly complex and less efficient.
This raises a critical question: Is it possible to formulate a Single Relaxation Time Lattice Boltzmann Method (SRT-LBM) that achieves the stability typically associated with the MRT approach? This study focuses on addressing this question.
Building on the multiple-relaxation-time lattice Boltzmann method (MRT-LBM) \cite{Lallemand.etc:2000}, we propose a new single MRT-featured Lattice Boltzmann Method (SmrtLBM) as follows
\begin{eqnarray}
	&& \hspace{-1cm}
	f_\alpha(x_j + e_{\alpha j} \delta t, t + \delta t) 
	=  f_\alpha^{eq}(x_j, t)  +
	(-1)^\alpha \left (  1  - \frac{1}{\tau}  \right )  \times \nonumber  \\
	&&  \hspace{1cm} 
  \sum_{k=1}^8  	\frac{(-1)^k}{4} 
  \left [ 	\mathbf{1}_{\{1, 2, 3, 4\}}(\alpha, k)   +  \mathbf{1}_{\{5, 6, 7, 8\}}(\alpha, k) \right ]
   \left [ f_k(x_j, t) -f_k^{eq}(x_j, t) \right ],   
	\label{SmrtLBM-1} 
\end{eqnarray}
where $\mathbf{1}_{\{a, b, c, d\}}(m,n)$ denotes the   indicator function defined as
\begin{equation}
\mathbf{1}_{\{a, b, c, d\}}(m, n) =
\begin{cases}
	1, & \text{if } m \in \{a, b, c, d\} \text{ and } n  \in \{a, b, c, d\}, \\
	0, & \text{otherwise}.
\end{cases}
\label{indicator-fn}
\end{equation}
The development of the SmrtLBM is detailed in the Appendix. Essentially, it builds upon the classic MRT-LBM framework by simplifying the relaxation times: all are fixed at 1, except for the two linked to kinematic viscosity. These two specific relaxation times are set to be equal and determined along with the time step and the lattice size to achieve the desired flow viscosity through \eq{relation-chapman-viscosity}. This approach ensures that the SmRT-LBM retains the inherent stability characteristics of the MRT-LBM.
The major advantage of the SmrtLBM lies in its simplicity, as it employs a single relaxation time and follows a procedure that is effectively as straightforward as that of the standard SRT-LBM. After the distribution function $f_\alpha$ is obtained using \eq{SmrtLBM-1}, the flow velocity and density are calculated using \eq{fea-0}.

To validate the SmrtLBM, numerical simulations of cavity flow in a square domain with dimensions of $1 \times 1$ in the horizontal ($x$) and vertical ($y$) directions are conducted at various Reynolds numbers. 
Cavity flow is a well-known complex flow characterised by the presence of vortices of different scales within a simple square domain. No-slip boundary conditions ($u_x=0$, $u_y=0$) are imposed on three fixed sides of the cavity, except for the top lid, where the boundary conditions are specified as $u_x=u_0$ and $u_y=0$, with $u_0=1$. The Reynolds number is defined as $R_e = u_0/\nu$. In these simulations, the standard bounce-back scheme is employed for the no-slip boundary conditions, while a relaxation time of $\tau = 0.51$ is used unless otherwise stated.

\begin{figure}[H]%
	\centering
	\subfloat[\centering Profile comparisons of $u_x(y)$ and  $u_y(x)$.]{{\includegraphics[width=8.7cm]{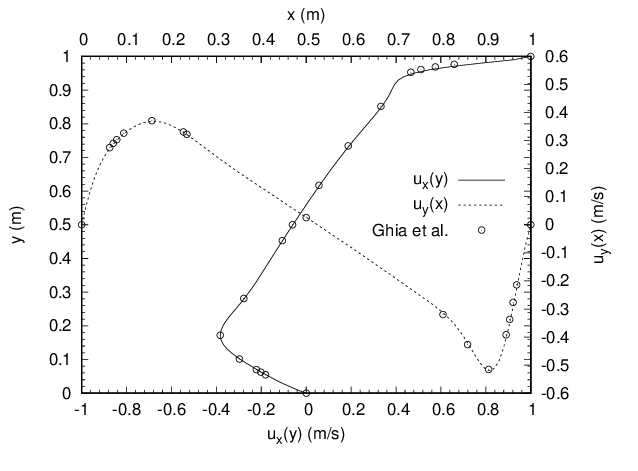} }}%
	\subfloat[\centering Streamlines.]{\raisebox{0.7cm}{\includegraphics[width=7.7cm]{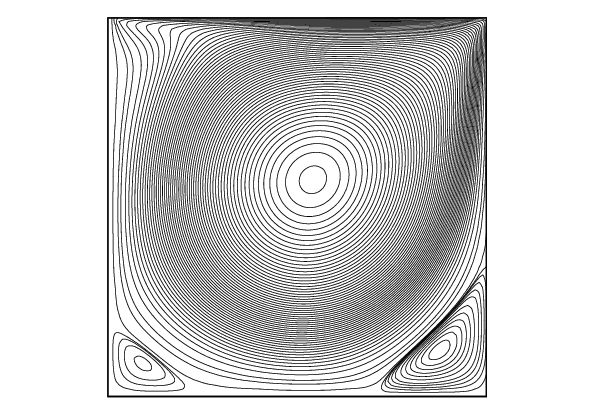} }  }%
	\caption{2D cavity flow for $R_e = 1000$. (a) Velocity profiles of  $u_x(y)$ and  $u_y(x)$ through the geometric centre of the cavity are compared with the numerical solution by Ghia et al. \cite{Ghia.etc:1982}, and (b) 
		the flow pattern in streamline shows a primary vortex and two secondary vortices.}%
	\label{cavity-Re1000}%
\end{figure}

First of all, we use a $100 \times 100$ lattice to simulate the cavity flow at $R_e = 1000$. A steady-state solution was achieved after the 60,000$^{th}$ time step. The numerical results are then compared with the well-established solution by Ghia et al. \cite{Ghia.etc:1982} and plotted in Fig.~\ref{cavity-Re1000}(a). The comparison clearly demonstrates that the SmrtLBM accurately reproduces the correct solution.  The flow pattern in streamline shows a primary vortex and two secondary vortices in Fig.~\ref{cavity-Re1000}(b).

\begin{figure}[H]%
	\centering
	\subfloat[\centering Profile comparisons of $u_x(y)$ and  $u_y(x)$.]{{\includegraphics[width=8.7cm]{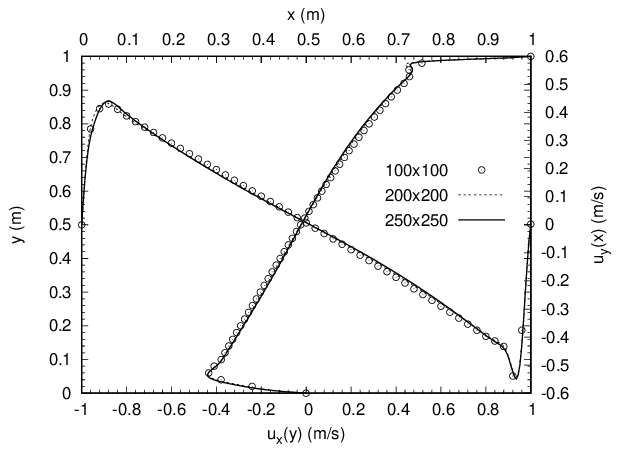}}}%
		\subfloat[\centering Velocity vectors.]{\raisebox{0.6cm}{\includegraphics[width=7.3cm]{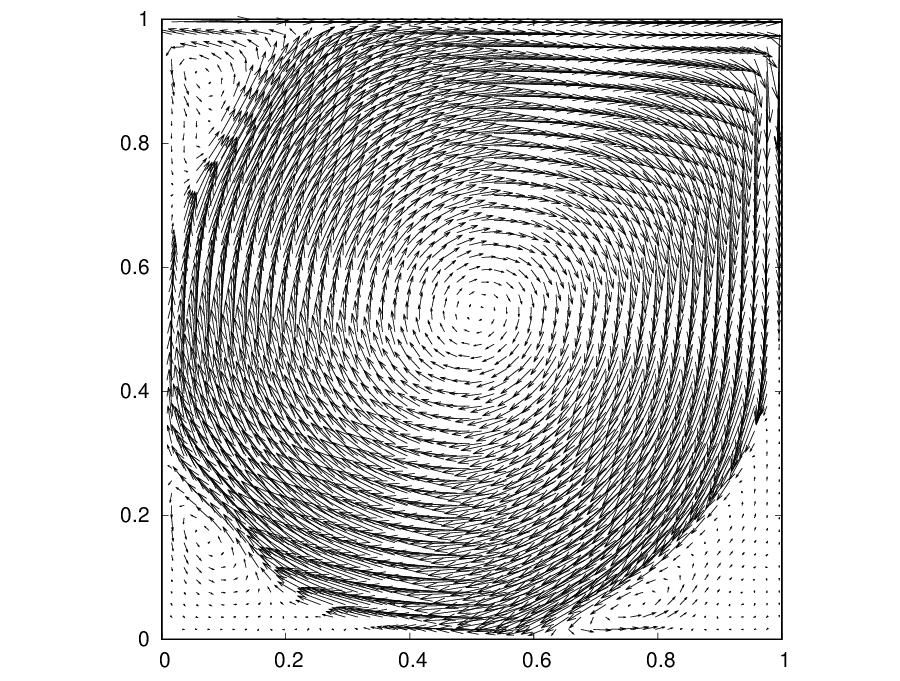} }}%
	\caption{2D cavity flow for $R_e = 10,000$ using three lattice numbers of $100 \times 100$, $200 \times 200$ and $250 \times 250$. (a) Velocity profiles of  $u_x(y)$ and  $u_y(x)$ through the geometric centre of the cavity are compared across three lattices, and 	(b) the flow pattern in velocity vectors using a $250 \times 250$ lattice shows a primary vortex and five secondary vortices.}%
	\label{cavity-Re10k-latticeCom}%
\end{figure}

\begin{figure}[H]%
	\centering
	\subfloat[\centering Profile comparisons of $u_x(y)$ and  $u_y(x)$.]{{\includegraphics[width=8.7cm]{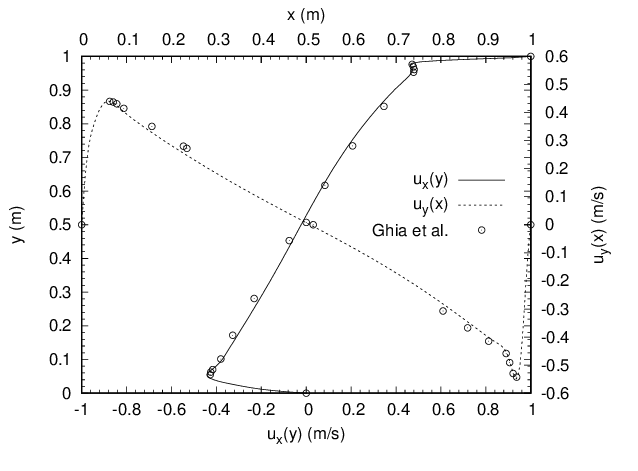} }}%
	\subfloat[\centering Streamlines.]{\raisebox{0.7cm}{\includegraphics[width=7.7cm]{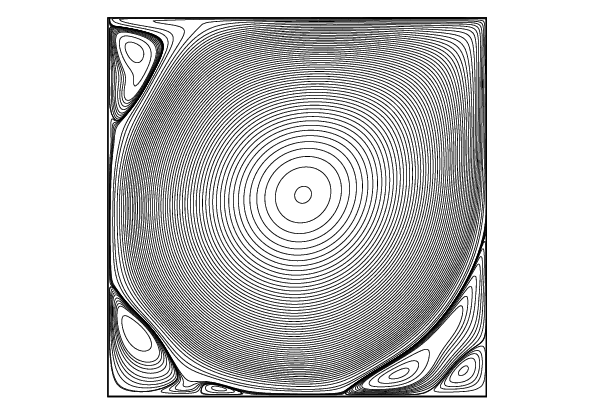} }}%
	\caption{2D cavity flow for $R_e = 10,000$. The steady solution using a $250 \times 250$ lattice is further compared and depicted for accuracy and flow features. (a) Velocity profiles of  $u_x(y)$ and  $u_y(x)$ through the geometric centre of the cavity are compared with the numerical solution by Ghia et al. \cite{Ghia.etc:1982}, and (b) the streamline flow pattern reveals a primary vortex along with five additional vortices using a $250 \times 250$ lattice.}
	\label{Re10k-250x250-stream}%
\end{figure}

\vspace{-0.5cm}
Next, to evaluate stability and accuracy, we simulate the cavity flow at a higher Reynolds number $R_e=10,000$, using three lattice resolutions: $100 \times 100$, $200 \times 200$ and $250 \times 250$. All other parameters remains unchanged. The velocity profiles of  $u_x(y)$ and  $u_y(x)$ through the geometric centre of the cavity are compared in Fig.~\ref{cavity-Re10k-latticeCom}(a). The velocity vectors are shown in Fig.~\ref{cavity-Re10k-latticeCom}(b). 
 After comparison with the numerical solution by Ghia et al.\cite{Ghia.etc:1982}, the results based on $250 \times 250$ lattices produce the accurate solutions as shown in Fig~\ref{Re10k-250x250-stream}(a). The flow pattern in streamlines is depicted in Fig~\ref{Re10k-250x250-stream}(b), showing a primary vortex and five secondary vortices.  It is noteworthy that SRT-LBM is capable of simulating 2D cavity flow up to a maximum Reynolds number of 
$R_e=7,500$  on a $256 \times 256$ lattice for a steady-state solution  \cite{Hou.etc:1995}.

\begin{figure}[H]%
	\centering
	\subfloat[\centering Velocity vectors at  208.3 s.]{{\includegraphics[width=7.7cm]{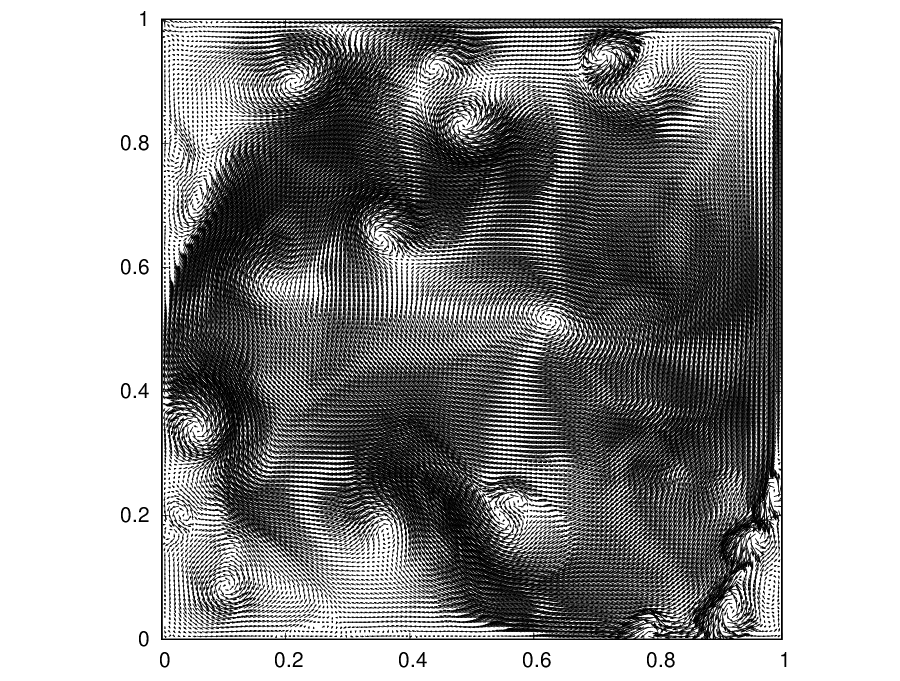} }}%
	\hspace{1.1cm}
	\subfloat[\centering Velocity vectors  at 416.6 s.]{{\includegraphics[width=7.7cm]{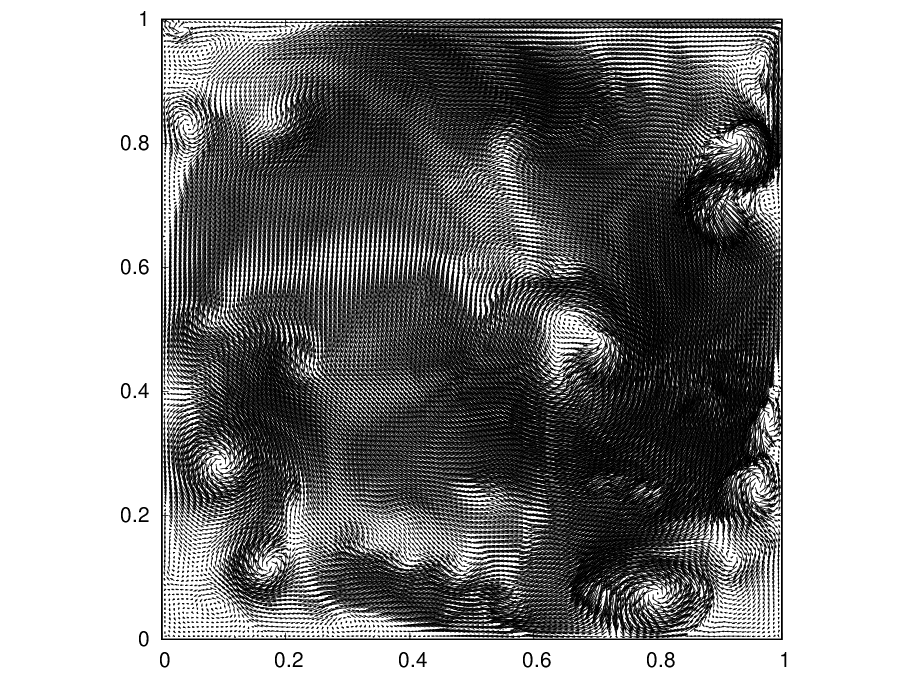} }}%
	\caption{2D cavity flow for $R_e = 100,000$ using $400 \times 400$ lattices. As there is no steady solution, two instantaneous velocity vectors are plotted, showing more than ten vortices in the flow.}%
	\label{result-Re100k-t502-400x400-1000k_vec}%
\end{figure}

Finally, we simulated the flow at $R_e=100,000$ using  a $400 \times 400$ lattice with $\tau = 0.502$ to further assess the stability of the model. 
The numerical simulations indicate that while the fully developed turbulent flow is achieved after 100 seconds, the flow remains highly unstable and does not reach a steady state.
The current simulation was run for over 400 seconds, corresponding to 100,000 time steps, without any instability issues arising. This suggests that the model remains stable even for very high Reynolds number or nearly negligible viscosity flows. The two instantaneous velocity fields, depicted in Fig.~\ref{result-Re100k-t502-400x400-1000k_vec}, reveal the presence of more than ten vortices. 

In conclusion, the SmrtLBM model emerges as a highly robust and reliable tool for fluid flow simulation, effectively balancing the stability of the complex MRT approach with the streamlined efficiency of a single relaxation time method. This makes it particularly well-suited for addressing complex flow challenges in both research and engineering. Its simplicity and efficiency are not only advantageous but also highly adaptable, allowing for straightforward extensions to other lattice Boltzmann models, such as those developed for shallow water flows \cite{zhoueLABSWE:2011} and axisymmetric flows \cite{zhouAxLABR:2011}. This adaptability significantly enhances the model's applicability, paving the way for new avenues of scientific exploration and innovation across various fields of science and engineering through a simple yet powerful lattice Boltzmann approach.



\titleformat{\section}{\normalfont\Large\bfseries}{Appendix}{1em}{}
\appendix

\section{Derivation}
\label{Methods}

We provide a detailed derivation of SmrtLBM. To begin, we review the conventional MRT lattice Boltzmann equation, which is
\begin{eqnarray}
	{\bf f} ({\bf x} + {\bf e} \delta t, t + \ \delta t) 
	= 
	{\bf f} ({\bf x}, t)   
	- \
	\mathbf{M}^{-1} \mathbf{S}   \mathbf{M} (\mathbf{f}- \mathbf{f}^{eq}),
	\label{lb.1} 
\end{eqnarray}
where ${\bf f} = [f_0,f_1,f_2,f_3,f_4,f_5,f_6,f_7,f_8]^T$ is the column vectors of the nine distribution functions of particles; ${\bf f}^{eq} = [f_0^{eq},f_1^{eq},f_2^{eq},f_3^{eq},f_4^{eq},f_5^{eq},f_6^{eq},f_7^{eq},f_8^{eq}]^T$ represents the column vectors of the nine local equilibrium distribution functions defined in \eq{feq-full};
$\mathbf{S}=\mbox{diag}[S_0,S_1,S_2,S_3,S_4,S_5,S_6,S_7,S_8]$ is a diagonal matrix of relaxation times; and $\mathbf{M}^{-1}$ is the inverse of the transformation matrix $\mathbf{M}$ as specified by
\begin{equation}
	\mathbf{M}=
	\left[ 
	\begin{array}{rrrrrrrrr}
	     1   &  1 &  1 &  1 &  1  &  1 &  1  &  1  &  1 \\
		-4 & -1 & -1 & -1 & -1 & 2 & 2 & 2 & 2 \\
		4 & -2 & -2 & -2 & -2 & 1 & 1 & 1 & 1 \\
		0 & 1 & 0 & -1 & 0 & 1 & -1 & -1 & 1 \\
		0 & -2 & 0 & 2 & 0 & 1 & -1 & -1 & 1 \\
		0 & 0 & 1 & 0 & -1 & 1 & 1 & -1 & -1 \\
		0 & 0 & -2 & 0 & 2 & 1 & 1 & -1 & -1 \\
		0 & 1 & -1 & 1 & -1 & 0 & 0 & 0 & 0 \\
		0 & 0 & 0 & 0 & 0 & 1 & -1 & 1 & -1 
	\end{array} 
	\right ],
	\label{M-matrix} 
\end{equation}
If we assume that $S_0=S_1=S_2=S_3=S_4=S_5=S_6=1/\tau_s$ and $S_7=S_8=1/\tau$,  	$\mathbf{S}$ can be expressed as the sum of two diagonal matrices below
\begin{equation}
	\mathbf{S}=
	\begin{bmatrix}
		1/\tau_s & 0 & 0 & 0 & 0 & 0 & 0 & 0 & 0 \\
		0 & 1/\tau_s & 0 & 0 & 0 & 0 & 0 & 0 & 0 \\
		0 & 0 & 1/\tau_s & 0 & 0 & 0 & 0 & 0 & 0 \\
		0 & 0 & 0 & 1/\tau_s & 0 & 0 & 0 & 0 & 0 \\
		0 & 0 & 0 & 0 & 1/\tau_s & 0 & 0 & 0 & 0 \\
		0 & 0 & 0 & 0 & 0 & 1/\tau_s & 0 & 0 & 0 \\
		0 & 0 & 0 & 0 & 0 & 0 & 1/\tau_s & 0 & 0 \\
		0 & 0 & 0 & 0 & 0 & 0 & 0 & 1/\tau & 0 \\
		0 & 0 & 0 & 0 & 0 & 0 & 0 & 0 & 1/\tau \\
	\end{bmatrix}
	=\mathbf{S}_t+\mathbf{S}_0
	\label{S-matrix:1} 
\end{equation}
in which
\begin{equation}
	\mathbf{S}_t=\frac{1}{\tau_s}
	\begin{bmatrix}
		1 & 0 & 0 & 0 & 0 & 0 & 0 & 0 & 0 \\
		0 & 1 & 0 & 0 & 0 & 0 & 0 & 0 & 0 \\
		0 & 0 & 1 & 0 & 0 & 0 & 0 & 0 & 0 \\
		0 & 0 & 0 & 1 & 0 & 0 & 0 & 0 & 0 \\
		0 & 0 & 0 & 0 & 1 & 0 & 0 & 0 & 0 \\
		0 & 0 & 0 & 0 & 0 & 1 & 0 & 0 & 0 \\
		0 & 0 & 0 & 0 & 0 & 0 & 1 & 0 & 0 \\
		0 & 0 & 0 & 0 & 0 & 0 & 0 & 1 & 0 \\
		0 & 0 & 0 & 0 & 0 & 0 & 0 & 0 & 1 \\
	\end{bmatrix}
	\label{St-matrix:1} 
\end{equation}
and
\begin{equation}
	\mathbf{S}_0=\left ( \frac{1}{\tau}-\frac{1}{\tau_s} \right )
	\begin{bmatrix}
		0 & 0 & 0 & 0 & 0 & 0 & 0 & 0 & 0 \\
		0 & 0 & 0 & 0 & 0 & 0 & 0 & 0 & 0 \\
		0 & 0 & 0 & 0 & 0 & 0 & 0 & 0 & 0 \\
		0 & 0 & 0 & 0 & 0 & 0 & 0 & 0 & 0 \\
		0 & 0 & 0 & 0 & 0 & 0 & 0 & 0 & 0 \\
		0 & 0 & 0 & 0 & 0 & 0 & 0 & 0 & 0 \\
		0 & 0 & 0 & 0 & 0 & 0 & 0 & 0 & 0 \\
		0 & 0 & 0 & 0 & 0 & 0 & 0 & 1 & 0 \\
		0 & 0 & 0 & 0 & 0 & 0 & 0 & 0 & 1 \\
	\end{bmatrix}
	\label{S0-matrix:1} 
\end{equation}
Substitution of \eq{S-matrix:1}  into \eq{lb.1}  results in
\begin{eqnarray}
	{\bf f} ({\bf x} + {\bf e} \delta t, t + \ \delta t) 
	= 
	{\bf f} ({\bf x}, t)   
	- \
	\mathbf{M}^{-1} (\mathbf{S}_t+ \mathbf{S}_0)  \mathbf{M} (\mathbf{f}- \mathbf{f}^{eq}),
	\label{lb.1.1} 
\end{eqnarray}
which can be simplified as
\begin{eqnarray}
	{\bf f} ({\bf x} + {\bf e} \delta t, t + \ \delta t) 
	= 
	{\bf f} ({\bf x}, t)   	- \frac{1}{\tau_s} (\mathbf{f}- \mathbf{f}^{eq})
	- \
	\mathbf{M}^{-1} \mathbf{S}_0  \mathbf{M} (\mathbf{f}- \mathbf{f}^{eq}).
	\label{lb.1.2} 
\end{eqnarray}
Evaluation of $\mathbf{M}^{-1} \mathbf{S}_0  \mathbf{M} (\mathbf{f}- \mathbf{f}^{eq}) $ yields
\renewcommand{\arraystretch}{1.6}
\begin{equation}
	\mathbf{M}^{-1} \mathbf{S}_0  \mathbf{M} (\mathbf{f}- \mathbf{f}^{eq}) 
	= \left (  \frac{1}{\tau_s} -\frac{1}{\tau} \right )
	\begin{bmatrix}
		0 \\
		-	\frac{1}{4} [(f_1-f_1^{eq}) - (f_2-f_2^{eq})+ (f_3-f_3^{eq}) - (f_4-f_4^{eq})]  \\
		\ \	 \frac{1}{4} [(f_1-f_1^{eq}) - (f_2-f_2^{eq})+ (f_3-f_3^{eq}) - (f_4-f_4^{eq})]  \\
		-	\frac{1}{4} [(f_1-f_1^{eq}) - (f_2-f_2^{eq})+ (f_3-f_3^{eq}) - (f_4-f_4^{eq})]  \\
		\ \	 \frac{1}{4} [(f_1-f_1^{eq}) - (f_2-f_2^{eq})+ (f_3-f_3^{eq}) - (f_4-f_4^{eq})]  \\
		-	\frac{1}{4} [(f_5-f_5^{eq}) - (f_6-f_6^{eq})+ (f_7-f_7^{eq}) - (f_8-f_8^{eq})]  \\
		\ \	\frac{1}{4} [(f_5-f_5^{eq}) - (f_6-f_6^{eq})+ (f_7-f_7^{eq}) - (f_8-f_8^{eq})]  \\
		-	\frac{1}{4} [(f_5-f_5^{eq}) - (f_6-f_6^{eq})+ (f_7-f_7^{eq}) - (f_8-f_8^{eq})]  \\
		\ \	\frac{1}{4} [(f_5-f_5^{eq}) - (f_6-f_6^{eq})+ (f_7-f_7^{eq}) - (f_8-f_8^{eq})] \\	
	\end{bmatrix}.
	\label{M-SM_ffeq-matrix:1} 
\end{equation}
If setting  
\begin{equation}
C_1 = \sum_{k=1}^4  \frac{(-1)^k}{4}  \left [f_k(x_j, t) -f_k^{eq}(x_j, t) \right ],
	 \hspace{10mm}
C_2 =  \sum_{k=5}^8   \frac{(-1)^k}{4} \left [f_k(x_j, t) -f_k^{eq}(x_j, t) \right ],
	\label{sigma1-2-0}
\end{equation}
we have
\begin{equation}
	\mathbf{M}^{-1} \mathbf{S}_0  \mathbf{M} (\mathbf{f}- \mathbf{f}^{eq}) =   \left ( \frac{1}{\tau_s} - \frac{1}{\tau}  \right ) [0, C_1, -C_1, C_1, -C_1, C_2, -C_2, C_2, -C_2]^T = \left ( \frac{1}{\tau_s} - \frac{1}{\tau} \right ) \boldsymbol{\sigma},
	\label{M-SM_ffeq-matrix:2} 
\end{equation}
in which 
\begin{equation}
\boldsymbol{\sigma} =  [0, C_1, -C_1, C_1, -C_1,C_2, -C_2,C_2, -C_2]^T. 
	\label{M-SM_sgma} 
\end{equation}
Substituting \eq{M-SM_ffeq-matrix:2}  into \eq{lb.1.2}  leads to
\begin{eqnarray}
	{\bf f} ({\bf x} + {\bf e} \delta t, t + \ \delta t) 
	= 
	{\bf f} ({\bf x}, t)   	- \frac{1}{\tau_s} (\mathbf{f}- \mathbf{f}^{eq})
	-   \left ( \frac{1}{\tau_s} - \frac{1}{\tau} \right )	\boldsymbol{\sigma},
	\label{lb.1.3a0} 
\end{eqnarray}
which can be expressed in component form or index notation for the distribution function $f_\alpha$,
\begin{eqnarray}
	&& \hspace{-0.85cm}
	f_\alpha(x_j + e_{\alpha j} \delta t, t + \delta t) 
	=  f_\alpha(x_j, t)  - \frac{1}{\tau_s} (f_\alpha- f_\alpha^{eq}) 
	-\left ( \frac{1}{\tau_s} - \frac{1}{\tau} \right )	\sigma_\alpha
	\label{lb.1.3a} 
\end{eqnarray}
By applying \eqs{indicator-fn}, substitution of \eq{M-SM_sgma} into above equation results in
\begin{eqnarray}
	&& \hspace{-0.85cm}
f_\alpha(x_j + e_{\alpha j} \delta t, t + \delta t) 
=  f_\alpha(x_j, t)  - \frac{1}{\tau_s} (f_\alpha- f_\alpha^{eq})  \nonumber  \\
&&  \hspace{2.95cm} + \
(-1)^\alpha  \left ( \frac{1}{\tau_s} - \frac{1}{\tau} \right )
  \sum_{k=1}^4  \mathbf{1}_{\{1, 2, 3, 4\}}(\alpha, k) 
\frac{(-1)^k}{4}  \left [f_k(x_j, t) -f_k^{eq}(x_j, t) \right ] \nonumber  \\
&&  \hspace{2.95cm} + \
(-1)^\alpha \left ( \frac{1}{\tau_s} - \frac{1}{\tau} \right )
  \sum_{k=5}^8   \mathbf{1}_{\{5, 6, 7, 8\}}(\alpha, k)
\frac{(-1)^k}{4} 
\left [f_k(x_j, t) -f_k^{eq}(x_j, t) \right ],
\label{lb.1.4} 
\end{eqnarray}
which is further written as
\begin{eqnarray}
	&& \hspace{-1cm}
	f_\alpha(x_j + e_{\alpha j} \delta t, t + \delta t) 
	=  f_\alpha(x_j, t)  - \frac{1}{\tau_s} (f_\alpha- f_\alpha^{eq})  +
(-1)^\alpha 	 \left ( \frac{1}{\tau_s} - \frac{1}{\tau} \right ) \times \nonumber  \\
	&&  \hspace{1cm} 
	\sum_{k=1}^8  	\frac{(-1)^k}{4} 
	\left [ 	\mathbf{1}_{\{1, 2, 3, 4\}}(\alpha, k)   +  \mathbf{1}_{\{5, 6, 7, 8\}}(\alpha, k) \right ]
	\left [f_k(x_j, t) -f_k^{eq}(x_j, t) \right ].   
	\label{SmrtLBM-lb.1.5} 
\end{eqnarray}
The above equation  involves two relaxation times, $\tau_s$ and $\tau$.  Both can take any positive number larger than 0.5.  The parameter $\tau$ is is related to flow kinematic viscosity by
\begin{equation}
	\nu =\frac{1}{6} ( 2 \tau - 1 ) e \delta x.
	\label{relation-chapman-viscosity}
\end{equation}
\eq{SmrtLBM-lb.1.5} effectively represents a two-relaxation-time LBM, which may offer additional advantages for simulating complex flows, though the differences between it and the SmrtLBM are minimal in the numerical simulations presented here.

Notably, by setting $\tau_s = \tau$, \eq{SmrtLBM-lb.1.5} reduces to the standard SRT-LBM. When $\tau_s = 1$, it simplifies to the proposed single relaxation time lattice Boltzmann equation \eqn{SmrtLBM-1}. Since this method is derived from the MRT scheme, it retains the MRT's enhanced stability and accuracy, hence it is referred to as the Single MRT-featured Lattice Boltzmann Method (SmrtLBM).

The recovery of the Navier-Stokes equations from the SmrtLBM follows the same procedure as the standard MRT approach, which is well-documented in the literature, such as in \cite{zhou2012MRT}.

\bibliography{smrtlbm}

\end{document}